\documentclass[floatfix,showpacs,aps,nofootinbib,onecolumn]{revtex4}

\usepackage{amssymb}

\usepackage[dvips]{epsfig}
\usepackage[english]{babel}
\usepackage{verbatim}
\usepackage{bbm}
\usepackage[utf8]{inputenc}
\usepackage{array}
\usepackage{amsmath}
\usepackage{amsfonts}
\usepackage{amssymb}

\usepackage{hyperref}
\usepackage[dvipsnames]{xcolor}
\hypersetup{colorlinks=true,linkbordercolor=Blue,linkcolor=Blue, citecolor=Blue}

\usepackage{subeqnarray}

\usepackage{epstopdf}
\usepackage{tensor}
\usepackage{indentfirst} 

\makeatletter
\renewcommand*\l@section{\@dottedtocline{1}{1.5em}{2.3em}}
\makeatother
\RequirePackage[normalem]{ulem} 
\RequirePackage{color}\definecolor{RED}{rgb}{1,0,0}\definecolor{BLUE}{rgb}{0,0,1} 

\usepackage{placeins}
\usepackage{float}

\begin{document}

\title{An alternative approach concerning Elko spinors and the hidden unitarity}

\author{L. C. Duarte$^{1}$}\email{laura.duarte@feg.unesp.br}
\author{R. de C. Lima$^{2}$}\email{rodrigo.lima@feg.unesp.br}
\author{R. J. Bueno Rogerio$^{3}$}\email{rodolforogerio@unifei.edu.br}
\author{C. H. Coronado Villalobos$^{4}$}\email{carlos.coronado@inpe.br}

\affiliation{ \\$^{1,2}$Universidade Estadual Paulista (Unesp)\\Faculdade de Engenharia, Guaratinguet\'a, Departamento de F\'isica e Qu\'imica\\
12516-410, Guaratinguet\'a, SP, Brazil.}

\affiliation{$^{3}$Instituto de F\'isica e Qu\'imica, Universidade Federal de Itajub\'a - IFQ (UNIFEI), \\
Av. BPS 1303, CEP 37500-903, Itajub\'a - MG, Brasil.}
\affiliation{$^4$Instituto Nacional de Pesquisas Espaciais (INPE),
12227-010, S\~ao Jos\'e dos Campos, SP, Brazil.}






\begin{abstract}\noindent\rule{14.15cm}{0.5pt}\\
\textbf{Abstract.} In this theoretical communication we look towards understand the underlying phenomenology concerning the Elko spinors within VSR theory. The program to be accomplished here start when we define the eigenspinors of the charge conjugation operator as  eigenstates of the helicity operator in the Cartesian coordinates system. This prescription is very useful in the sense of phenomenological point of view, so, we propose a set of Elko spinors ready to be computationally implemented. Regardless of, in order to show the application of given approach we impose to these spinors to be restrict to an axis, coincidentally the axis of locality \cite{jcap,hep-ph}, and then, using the proposed prescription, we search for physical amounts and physical processes by analysing the Yukawa and the self-interaction in such framework.
\\
\noindent\rule{14.15cm}{0.5pt}\\
\end{abstract}


\maketitle

\maketitle
\section{Introduction}
Elko spinors are a recent theoretical set of a spin one-half fermionic objects, endowed with mass-dimension-one, dual helicity feature, exhibit neutral behaviour under charge conjugation operator. The dynamic of the Elko spinors field are the very same dynamic of the scalar field, they are governed exclusively by the Klein-Gordon equation \cite{AH1, ahluwaliadark}. Such spinor emerge in the literature as a natural candidate to describe dark matter due to renormalizable self-interaction and limited electromagnetic interactions. By the aforementioned reasons, it plays a protagonist role in several areas of physics and the scientific community look  towards unveil its nature and understand what composes dark matter. 

Until the year 2016, it was believed that Elko breaks the relativistic Lorentz covariance, with a breaking term encoded in the spin sums, thus, showing locality only  along $z$-axis \cite{jcap,somasdespin}. Then, the Lorentz invariance was firstly circumvented by the \emph{Very Special Relativity} (VSR), a theory proposed by Cohen and Glashow \cite{cohen}, which coincidentally leaves the spin sums invariant under $HOM(2)$ and $SIM(2)$ group transformations \cite{horvath, ahluwaliahorvath}. We remark the relevance of the VSR theory by paying attention on recent researches developed concerning phenomenology (and other subjects) within this context \cite{alfaro1, alfaro2, alfaro3, alfaro4, alfaro5, alfaro6, selva, bufalo, alekha, alekha2}.  

For a long time it was believed that the non-locality feature was regarded as an inherent characteristic of dark matter. Recently, in \cite{AH1} it was proposed a subtle deformation on the Elko adjoint spinor, leading to a \emph{new local field},\footnote{It is important to note the conceptual difference between Elko spinors (VSR invariant) and the \emph{new local fields} (Lorentz invariant). Henceforth, we will make use of the correct nomenclature to distinguish them. For a better understanding, authors recommend Ref.\cite{AH1}.} and for that reason the problem of spin sum do not show invariance (or covariance) under Lorentz transformations was bypassed. We emphasize that we strongly agree with the recent mass-dimension-one formulation, however, our focus is to explore the previous formulation and consequently better understand both scenarios. We elucidate to the reader about the existent branch between the recent and the previous formulation. In this context, it is clear that the new formulation stands for mass-dimension-one spinors, which belong to Lorentz group, making necessary a deep review in the Weinberg's no-go theorem \cite{nogo}, while the previous one is restricted to the symmetries of VSR theory, so, both formulations are distinct. 

The quest to place Elko spinors at a well-established level is constant. As regards the papers \cite{diracelko, diracelkoaction}, researches concerning a better understanding of mass-dimension-one spinors, taking into account the well-established mass dimension $3/2$ spinors, still being focus in several areas. In the case of mapping Elko into Dirac fields was recently developed in \cite{tipo4} where the authors maps Elko into single helicity spinors and as a net result they obtain Dirac spinors and also is possible to obtain type-4 spinors, explicitly. Regarding to the bilinear analysis, we should emphasize that Elko spinors can not fit within Lounesto classification (although, naively, it can be labelled as a type-5), but it is not totally correct due to the contrasting dual structure that Elko carry. So, it is necessary to develop a new classification for mass-dimension-one fermions or a general classification where it takes into account any dual structure. 
Cohen and Glashow states that the VSR theory implies in special relativity either in the context of local quantum field theory or in the framework of $CP$ conservation \cite{cohen}, merging VSR with Elko is still feasible \cite{cavalcantielko}. The amplitude for a two body decay of a spinless particle at rest may depend on the direction of the decay products relative to the VSR preferred direction \cite{cohen, cavalcantielko}, and VSR signature is expected to arise for the mass dimension one Fermi field \cite{elkosignature}.

In the present work, we firstly propose to write the Elko spinors in the Cartesian coordinate system, giving all the additional support to the computational implementation programs like \emph{FeynRules, Madgraph5}, among others, often used by phenomenologists, as commonly made for the other spinors present in the literature, e.g., Dirac, Majorana and Weyl spinors. Thereafter, in order to exhibit the usefulness of such prescription, we privilege a direction aiming to better understand the eminent physics encoded behind the axis of the locality.

From the phenomenological point of view, the relevant physical amounts use to come from the data obtained on the particle colliders, e.g., the decay-rate time and cross-section. In the QFT context, the most general property concerning scattering matrix ($\mathcal{S}$-matrix) is the unitarity. Given property guarantee, during a physical process, that the orthonormality relation in the asymptotic states must to be preserved. In general grounds, \emph{unitarity} simply express the conservation of the transition probability between two different states. It's worth pointing out that $\mathcal{S}$-matrix for non-local theories still not yet well established. Nevertheless, in the scope of the present work, which deals with the physics behind an object forced to be along the locality axis, all the above mentioned prescription must remain valid. Note that for the analysis of the phenomenological processes, both formulations along such axis must be indistinguishable. 

Until the present moment it was believed that all the Elko re-normalizable interactions violates the tree level unitarity, except the Yukawa interaction \cite{elko5}. We also analyse the cross-section for self-interaction and for Elko-Higgs interaction along $z$-axis, and in this specific framework both processes preserve the tree level unitarity. Therefore, the purpose of this review is to better understand the physical information encoded on the VSR theory.

This paper is organized as it follows: Section \ref{exordium} we start the program by defining the eigenstates of the helicity operator in the Cartesian coordinate system. In Section \ref{aplications} we define the Elko spinors in the Cartesian coordinate system. Section \ref{phenom} is reserved to deal with phenomenological analysis. Finally, in Section \ref{conclusions} we conclude.

\section{Exordium: Defining the eigenstates of helicity operator in the Cartesian coordinate system}\label{exordium}
We start this communication writting the helicity operator \cite{helicidade}, ${\Sigma}$, in the Cartesian coordinate system, as it follows
\begin{equation}\label{1}
 \Sigma \equiv \boldsymbol{\sigma}\cdot\hat{p},
\end{equation}
where $\boldsymbol{\sigma}$ stands for the Pauli matrices, given by 
\begin{eqnarray}
\sigma_{x} =  \begin{pmatrix}
                        0 & 1\\ 
                        1 & 0   
                       \end{pmatrix},\quad \sigma_{y} =  \begin{pmatrix}
                                                        0 & -i\\ 
                                                        i & 0   
                                                        \end{pmatrix} \;\mbox{and}\;\sigma_{z} =  \begin{pmatrix}
                                                                                         1 & 0\\ 
                                                                                         0 & -1   
                                                                                         \end{pmatrix}.
\end{eqnarray}                                                                                                                                                                     
In the momentum Cartesian coordinate system (${\boldsymbol{p} = p_{x}, p_{y}, p_{z}}$) the helicity operator reads
\begin{equation}\label{2}
 \Sigma = \boldsymbol{\sigma}\cdot\frac{\boldsymbol{p}}{|\boldsymbol{p}|} = \frac{1}{|\boldsymbol{p}|}\begin{pmatrix}
                                                                           p_{z}          & p_{x} - ip_{y}\\ 
                                                                           p_{x} + ip_{y} & -p_{z}   
                                                                          \end{pmatrix},
\end{equation}
where ${|\boldsymbol{p}| = \sqrt{p_{x}^{2} + p_{y}^{2} + p_{z}^{2}}}$.  The normalized eigenstates ${|+\rangle}$ and ${|-\rangle}$ form a complete set of eigenstates of the helicity operator with eigenvalues ${+1}$ and ${-1}$ respectively, which allow one to write
\begin{equation}\label{4}
 |+\rangle = \frac{1}{|\boldsymbol{p}|}                                                      \begin{pmatrix}
                                                                           p_{z}          & p_{x} - ip_{y}\\ 
                                                                           p_{x} + ip_{y} & -p_{z}   
                                                                          \end{pmatrix}\begin{pmatrix}
                                                                                        a          \\ 
                                                                                        b    
                                                                                        \end{pmatrix} = + \begin{pmatrix}
                                                                                                          a          \\ 
                                                                                                          b    
                                                                                                          \end{pmatrix}  ,
\end{equation}
and
\begin{equation}\label{5}
|-\rangle = \frac{1}{|\boldsymbol{p}|}                                                      \begin{pmatrix}
                                                                           p_{z}          & p_{x} - ip_{y}\\ 
                                                                           p_{x} + ip_{y} & -p_{z}   
                                                                          \end{pmatrix}\begin{pmatrix}
                                                                                        c          \\ 
                                                                                        d    
                                                                                        \end{pmatrix} = - \begin{pmatrix}
                                                                                                          c          \\ 
                                                                                                          d    
                                                                                                          \end{pmatrix} .
\end{equation}
Equations (\ref{4}) and (\ref{5}) provide the value of ${a}$, ${b}$, ${c}$ and ${d}$, in such a way
\begin{eqnarray}\label{6}
  a &=& \pm \frac{(p_{x} - ip_{y})}{\sqrt{2|\boldsymbol{p}|(|\boldsymbol{p}|-p_{z})}},\\
  b &=& \pm \sqrt{\frac{(|\boldsymbol{p}| - p_{z})}{2|\boldsymbol{p}|}},\\
  c &=& \mp \frac{(p_{x} - ip_{y})}{\sqrt{2|\boldsymbol{p}|(|\boldsymbol{p}| + p_{z})}},\\
  d &=& \pm \sqrt{\frac{(|\boldsymbol{p}| + p_{z})}{2|\boldsymbol{p}|}},
\end{eqnarray}
thus, the eigenstates $\vert + \rangle$ and  $\vert - \rangle$ in Cartesian coordinate system read
\begin{equation}\label{7}
 |+\rangle_{\pm} = \pm\frac{1}{\sqrt{2|\boldsymbol{p}|}}\begin{pmatrix}
                                            \frac{(p_{x} - ip_{y})}{\sqrt{|\boldsymbol{p}| - p_{z}}}\\ 
                                            \sqrt{|\boldsymbol{p}| - p_{z}}    
                                            \end{pmatrix},
\end{equation}
\begin{equation}\label{8}
 |-\rangle _{\pm}= \pm\frac{1}{\sqrt{2|\boldsymbol{p}|}}\begin{pmatrix}
                                            \frac{(-p_{x} + ip_{y})}{\sqrt{|\boldsymbol{p}| + p_{z}}}\\ 
                                            \sqrt{|\boldsymbol{p}| + p_{z}}    
                                            \end{pmatrix}.
\end{equation}
According to Ref.\cite{jcap}, the formal structure of the Elko spinor in the rest frame reference is composed by 
\begin{eqnarray}\label{19}
 \lambda^{S}_{\{-,+\}}(\boldsymbol{0}) &=&  \begin{pmatrix}
                               i\Theta[\phi_{L}^{+}(\boldsymbol{0})]^{*}\\ 
                               \phi_{L}^{+}(\boldsymbol{0})    
                               \end{pmatrix},
\\
 \lambda^{S}_{\{+,-\}}(\boldsymbol{0}) &=&  \begin{pmatrix}
                               i\Theta[\phi_{L}^{-}(\boldsymbol{0})]^{*}\\ 
                               \phi_{L}^{-}(\boldsymbol{0})    
                               \end{pmatrix},
\\
\label{21} \lambda^{A}_{\{-,+\}}(\boldsymbol{0}) &=&  \begin{pmatrix}
                               -i\Theta[\phi_{L}^{+}(\boldsymbol{0})]^{*}\\ 
                               \phi_{L}^{+}(\boldsymbol{0})    
                               \end{pmatrix},
\\
\label{212} \lambda^{A}_{\{+,-\}}(\boldsymbol{0}) &=&  \begin{pmatrix}
                               -i\Theta[\phi_{L}^{-}(\boldsymbol{0})]^{*}\\ 
                               \phi_{L}^{-}(\boldsymbol{0})    
                               \end{pmatrix},
\end{eqnarray}
where the index $S$ stands for self-conjugated spinors, $A$ stands for the anti-self-conjugated spinors and $\Theta$ 
is the Wigner time-reversal operator, in the spin $1/2$ representation is given by
\begin{eqnarray}
\Theta= \begin{pmatrix}
0 & -1\\ 
1 & 0  
\end{pmatrix},
\end{eqnarray}
such operator act as follows
\begin{eqnarray}
\Theta\boldsymbol{\sigma}\Theta^{-1} = -\boldsymbol{\sigma}^{*},
\end{eqnarray}
unlike the Dirac spinors, where the representation-space are usually related by the parity symmetry \cite{speranca}, here we see that we abstain from the discrete symmetries and relate the representation-space via $\Theta$ operator.

The equations \eqref{7} and \eqref{8} are the building blocks to the proposed approach, due to the fact that it makes possible to write the components $\phi_{L}^{+}(\boldsymbol{0})$ and $ \phi_{L}^{-}(\boldsymbol{0})$, so, we have
\begin{equation}\label{9}
 \phi_{L}^{+}(\boldsymbol{0}) = \sqrt{m}e^{i\vartheta_{1}}|+\rangle
\end{equation}
and
\begin{equation}\label{10}
 \phi_{L}^{-}(\boldsymbol{0}) = \sqrt{m}e^{i\vartheta_{2}}|-\rangle,
\end{equation}
where ${\vartheta_{1}}$ e ${\vartheta_{2}}$ are arbitrary phases. The presence of the mass factor in \eqref{9} and \eqref{10} is related to the massless limit case. In such scenario, spinors at rest in the $(1/2,0)$ and $(0,1/2)$ representation-space must to vanish \cite{ahluwagold}. By consistency, the interaction amplitudes must to have the ``normalization factor'' $m^{j}$, where $j$ stands for the spin of the particle \cite{marinov}. 

Since the eigenstates in (\ref{7}) and (\ref{8}) have a degree of freedom encoded in the signals presented in the lower index ($\vert\pm\rangle_\pm$) we are able to write four different \emph{types} of massive left-hand spinors, as displayed below
\begin{eqnarray}\label{13}
\mbox{\emph{TYPE 1}:}  \quad
 [\phi_{L}^{+}(\boldsymbol{0})]_{+} &=&  \sqrt{\frac{m}{2|\boldsymbol{p}|}}\begin{pmatrix}
                                                                                                                                            \frac{(p_{x} - ip_{y})}{\sqrt{|\boldsymbol{p}| - p_{z}}}\\ 
                                                                                                                                            \sqrt{|\boldsymbol{p}| - p_{z}}                                                                                                                                             \end{pmatrix},
\\
\label{14} [\phi_{L}^{-}(\boldsymbol{0})]_{+} &=& \sqrt{\frac{m}{2|\boldsymbol{p}|}}\begin{pmatrix}                                                                                                                                            \frac{(-p_{x} + ip_{y})}{\sqrt{|\boldsymbol{p}| + p_{z}}}\\                                                                                                                                             \sqrt{|\boldsymbol{p}| + p_{z}}                                                                                                                                             \end{pmatrix},
\end{eqnarray}
\begin{eqnarray}\label{11}
\mbox{\emph{TYPE 2}:}\quad
[\phi_{L}^{+}(\boldsymbol{0})]_{+} &=& \sqrt{\frac{m}{2|\boldsymbol{p}|}}\begin{pmatrix}                                                                                                                                       \frac{(p_{x} - ip_{y})}{\sqrt{|\boldsymbol{p}| - p_{z}}}\\                                                                                                                                            \sqrt{|\boldsymbol{p}| - p_{z}}                                                                                                                                                \end{pmatrix},
\\
\label{12}  [\phi_{L}^{-}(\boldsymbol{0})]_{-} &=&  \sqrt{\frac{m}{2|\boldsymbol{p}|}}\begin{pmatrix}                                                                                                                                            \frac{(p_{x} - ip_{y})}{\sqrt{|\boldsymbol{p}| + p_{z}}}\\                                                                                                                                            -\sqrt{|\boldsymbol{p}| + p_{z}}                                                                                                                                            \end{pmatrix},
\end{eqnarray}
\begin{eqnarray}\label{15}
\mbox{\emph{TYPE 3}:}\quad
 [\phi_{L}^{+}(\boldsymbol{0})]_{-} &=& \sqrt{\frac{m}{2|\boldsymbol{p}|}}\begin{pmatrix}                                                                                                                                            \frac{(-p_{x} + ip_{y})}{\sqrt{|\boldsymbol{p}| - p_{z}}}\\                                                                                                                                             -\sqrt{|\boldsymbol{p}| - p_{z}}                                                                                                                                                \end{pmatrix},\\
\label{16} [\phi_{L}^{-}(\boldsymbol{0})]_{+}  &=& \sqrt{\frac{m}{2|\boldsymbol{p}|}}\begin{pmatrix}                                                                                                                                            \frac{(-p_{x} + ip_{y})}{\sqrt{|\boldsymbol{p}| + p_{z}}}\\                                                                                                                                             \sqrt{|\boldsymbol{p}| + p_{z}}                                                                                                                                            \end{pmatrix},
\end{eqnarray}
\begin{eqnarray}\label{17}
\mbox{\emph{TYPE 4}:}\quad 
 [\phi_{L}^{+}(\boldsymbol{0})]_{-} &=& \sqrt{\frac{m}{2|\boldsymbol{p}|}}\begin{pmatrix}                                                                                                                                           \frac{(-p_{x} + ip_{y})}{\sqrt{|\boldsymbol{p}| - p_{z}}}\\                                                                                                                                            -\sqrt{|\boldsymbol{p}| - p_{z}}                                                                                                                                            \end{pmatrix},\\                                                                                                                        
\label{18} [\phi_{L}^{-}(\boldsymbol{0})]_{-} &=&  \sqrt{\frac{m}{2|\boldsymbol{p}|}}\begin{pmatrix}                                                                                                                                            \frac{(p_{x} - ip_{y})}{\sqrt{|\boldsymbol{p}| + p_{z}}}\\                                                                                                                                             -\sqrt{|\boldsymbol{p}| + p_{z}}                                                                                                                                         \end{pmatrix}.
\end{eqnarray}
Inserting the left-hand components presented in \eqref{13} and \eqref{14} into \eqref{19}-\eqref{212}, then, we are able to write the following set of Elko spinors\footnote{Where we have defined the boost factors as
$\mathcal{B}_{\pm} = \sqrt{\frac{E + m}{2m}}\left(1 \pm \frac{|\boldsymbol{p}|}{E+m}\right)$.} 
\begin{eqnarray}\label{27}
\mbox{\emph{TYPE 1}:}\nonumber\\\;
 \lambda^{S}_{\{-,+\}}(\boldsymbol{p}) &=& \sqrt{\frac{m}{2|\boldsymbol{p}|}}\mathcal{B}_{-}\begin{pmatrix}
                                                              -i\sqrt{|\boldsymbol{p}| - p_{z}}\\ 
                                                              \frac{(-p_{y}+ip_{x})}{\sqrt{|\boldsymbol{p}| - p_{z}}}\\
                                                              \frac{(p_{x}-ip_{y})}{\sqrt{|\boldsymbol{p}| - p_{z}}}\\
                                                              \sqrt{|\boldsymbol{p}| - p_{z}}
                                                             \end{pmatrix},
\end{eqnarray}
\begin{eqnarray}
 \lambda^{S}_{\{+,-\}}(\boldsymbol{p}) &=& \sqrt{\frac{m}{2|\boldsymbol{p}|}}\mathcal{B}_{+}\begin{pmatrix}
                                                              -i\sqrt{|\boldsymbol{p}| + p_{z}}\\ 
                                                              \frac{(p_{y}-ip_{x})}{\sqrt{|\boldsymbol{p}| + p_{z}}}\\
                                                              \frac{(-p_{x}+ip_{y})}{\sqrt{|\boldsymbol{p}| + p_{z}}}\\
                                                              \sqrt{|\boldsymbol{p}| + p_{z}}
                                                             \end{pmatrix},
\end{eqnarray}
\begin{eqnarray}
\label{29} \lambda^{A}_{\{-,+\}}(\boldsymbol{p}) &=& \sqrt{\frac{m}{2|\boldsymbol{p}|}}\mathcal{B}_{-}\begin{pmatrix}
                                                              i\sqrt{|\boldsymbol{p}| - p_{z}}\\ 
                                                              \frac{(p_{y}-ip_{x})}{\sqrt{|\boldsymbol{p}| - p_{z}}}\\
                                                              \frac{(p_{x}-ip_{y})}{\sqrt{|\boldsymbol{p}| - p_{z}}}\\
                                                              \sqrt{|\boldsymbol{p}| - p_{z}}
                                                             \end{pmatrix},
\end{eqnarray}
\begin{eqnarray}
 \lambda^{A}_{\{+,-\}}(\boldsymbol{p}) &=& \sqrt{\frac{m}{2|\boldsymbol{p}|}}\mathcal{B}_{+}\begin{pmatrix}
                                                              i\sqrt{|\boldsymbol{p}| + p_{z}}\\ 
                                                              \frac{(-p_{y}+ip_{x})}{\sqrt{|\boldsymbol{p}| + p_{z}}}\\
                                                              \frac{(-p_{x}+ip_{y})}{\sqrt{|\boldsymbol{p}| + p_{z}}}\\
                                                              \sqrt{|\boldsymbol{p}| + p_{z}}
                                                             \end{pmatrix},
\end{eqnarray}
the spinors above are dual helicity spinors besides satisfying the charge conjugation relation. In the next sections we explore all the physical properties of these spinors. For particular reasons that will soon become clear, we focus on \emph{TYPE 1} spinors, however, the results holds the same for all the other \emph{types}. 

\section{Exploring the Elko field in the local axis}\label{aplications}
This section is reserved to extend the analysis and applications of the Elko spinors built up until now. In this way, have seen that, as previously mentioned, such prescription is very useful in the sense of phenomenological applications, so, we analyse such spinors along $z$-axis looking towards explore the relevant physical information encoded in such framework aiming to test the machinery applied until now. 


As previously mentioned, the focus of the present work is to write the Elko spinors along the $z$-axis, where it is believed to be local \cite{jcap, hep-ph, cheng2d}, overcoming the spin sums invariance under Lorentz transformations problem and also locality; clearly preventing a concrete physical interpretation. From this point of view, we pursue our analysis by taking $p_{\mu}=(p_{0}, 0, 0, p_z)$. It is possible to note that the spinors ${\lambda^{S/A}_{\{-,+\}}}$, presented in \eqref{27} and \eqref{29} need a careful attention, since they are undetermined at the limit 
\begin{equation}
\lim_{\vert\boldsymbol{p}\vert\rightarrow p_z}\frac{1}{\sqrt{\vert\boldsymbol{p}\vert-p_z}}.
\end{equation}
In order to circumvent this issue, we perform some mathematical manipulations, so, taking the limits ${p_{x}\longrightarrow0}$ and ${p_{y}\longrightarrow0}$, we easily obtain
\begin{equation}\label{39}
 \pm\frac{1}{\sqrt{2|\boldsymbol{p}|}}\begin{pmatrix}
                             \frac{( p_{x}- ip_{y})}{\sqrt{|\boldsymbol{p}| - p_{z}}}\\
                             \sqrt{|\boldsymbol{p}| - p_{z}}
                            \end{pmatrix} \rightarrow \begin{pmatrix}
                             \pm 1\\
                               0
                            \end{pmatrix},
\end{equation}
and
\begin{equation}\label{40}
\pm \frac{1}{\sqrt{2|\boldsymbol{p}|}}\begin{pmatrix}
                             \frac{( p_{y}- ip_{x})}{\sqrt{|\boldsymbol{p}| - p_{z}}}\\
                            \sqrt{|\boldsymbol{p}| - p_{z}}
                            \end{pmatrix} \rightarrow \begin{pmatrix}
                             \mp i\\
                                0
                            \end{pmatrix},
\end{equation}
finally, the Elko spinors along the $z$-axis can be expressed as\footnote{Where we have defined the boost factors along the $z$-axis as
$\mathcal{B}^{z}_{\pm} = \sqrt{\frac{E + m}{2m}}\left(1 \pm \frac{|\boldsymbol{p_z}|}{E+m}\right)$.}  
\begin{eqnarray}\label{41}
\mbox{\emph{TYPE 1}:}\; 
 \lambda^{S}_{\{-,+\}}(\boldsymbol{p_z}) &=& \sqrt{m}\mathcal{B}^{z}_{-}\begin{pmatrix}
                                             0\\
                                             i\\
                                             1\\
                                             0
                                           \end{pmatrix},
\end{eqnarray}
\begin{eqnarray}
 \lambda^{S}_{\{+,-\}}(\boldsymbol{p_z}) &=& \sqrt{m}\mathcal{B}^{z}_{+}\begin{pmatrix}
                                             i\\
                                             0\\
                                             0\\
                                            -1
                                           \end{pmatrix},
\end{eqnarray}
\begin{eqnarray}
\label{43} \lambda^{A}_{\{-,+\}}(\boldsymbol{p_z}) &=& \sqrt{m}\mathcal{B}^{z}_{-}\begin{pmatrix}
                                             0\\
                                            -i\\
                                             1\\
                                             0
                                           \end{pmatrix},
\end{eqnarray}
\begin{eqnarray}
 \lambda^{A}_{\{+,-\}}(\boldsymbol{p_z}) &=& \sqrt{m}\mathcal{B}^{z}_{+}\begin{pmatrix}
                                            -i\\
                                             0\\
                                             0\\
                                            -1
                                           \end{pmatrix}.
\end{eqnarray}
Such spinors, labeled by \emph{TYPE 1}, match to the one previously obtained in \cite{cheng2d} and, for such reason they will be focus in the present work. It is important to remark that even in the preferred direction given spinors still being Elko spinors. Thus, the dual structure is the same as already defined in \cite{AH1}, i.e.,
\begin{eqnarray}\label{duall}
\stackrel{\neg}{\lambda}_{\{\mp,\pm\}}^{S/A}(\boldsymbol{p}) &=& [\Xi(\boldsymbol{p})\lambda_{\{\mp,\pm\}}^{S/A}(\boldsymbol{p})]^{\dag}\gamma^{0},\nonumber
\\
&=&\pm i[\lambda_{\{\pm,\mp\}}^{S/A}(\boldsymbol{p})]^{\dag}\gamma^{0},
\end{eqnarray}
where the $\Xi(\boldsymbol{p})$ operator is defined in \cite{bilineares}. In the context of this communication, $\Xi(\boldsymbol{p})$ along the preferred direction reads
\begin{eqnarray*}\label{84}
 \Xi(\boldsymbol{p_z}) = \begin{pmatrix}
                  0          &  -\frac{i(E + |p_{z}|)}{m} &      0      &        0\\
              \frac{i(E - |p_{z}|) }{m}  &      0        &      0      &        0       \\
                       0     &      0        &      0      &    -\frac{i(E - |p_{z}|)}{m}  \\
                    0        &      0        &  \frac{i(E + |p_{z}|)}{m}&        0
                   \end{pmatrix},
\end{eqnarray*}
and the dual spinor is defined as
\begin{eqnarray*}\label{61}
\mbox{\emph{TYPE 1}:}\;
 \stackrel{\neg}{\lambda}_{\{-,+\}}^{S}(\boldsymbol{p_z}) &=& \sqrt{m}\mathcal{B}^{z}_{+}\begin{pmatrix}
                                             0 & -i & 1 & 0
                                           \end{pmatrix},
\\
\label{62} \stackrel{\neg}{\lambda}_{\{+,-\}}^{S}(\boldsymbol{p_z}) &=& \sqrt{m}\mathcal{B}^{z}_{-}\begin{pmatrix}
                                             -i & 0 & 0 & -1
                                           \end{pmatrix},
\\
\label{63} \stackrel{\neg}{\lambda}_{\{-,+\}}^{A}(\boldsymbol{p_z}) &=& \sqrt{m}\mathcal{B}^{z}_{+}\begin{pmatrix}
                                             0 & -i & -1 & 0
                                           \end{pmatrix},
\\
\label{64}
 \stackrel{\neg}{\lambda}_{\{+,-\}}^{A}(\boldsymbol{p_z})&=& \sqrt{m}\mathcal{B}^{z}_{-}\begin{pmatrix}
                                             -i & 0 & 0 & 1
                                           \end{pmatrix},
\end{eqnarray*}
as expected, the orthonormal relations remains unchanged\footnote{Regarding the physical observables (bilinear forms) given in \cite{lounesto}, the authors in \cite{roldaoo} classify Elko sinors as type-5. However the Elko norm \eqref{77}, are defined taking into account the Elko dual structure, so, all the physical amounts should carry the same dual structure rather than the Dirac one. In this vein, in  \cite{bilineares} authors perform a very same procedure to define the bilinear amounts as developed in \cite{lounesto}}.
\begin{eqnarray}\label{77}
\stackrel{\neg}{\lambda}_{\beta}\!\!\!\!\;^{S/A}(\boldsymbol{p_z})\lambda^{S/A}_{\beta^{'}}(\boldsymbol{p_z}) = \pm 2m\delta_{\beta\beta^{'}}.
\end{eqnarray}

Next step is to evaluate the spin sums, the importance of this physical quantity lies in the fact that it compose the core of the transition amplitude and scattering, so, we have 
\begin{eqnarray}\label{83}
 \sum_{\beta}\lambda^{S}_{\beta}(\boldsymbol{p_{z}})\stackrel{\neg}{\lambda}_{\beta}^{S}(\boldsymbol{p_{z}}) &=& m(\mathbbm{1}+G_z), 
\end{eqnarray}
and
\begin{eqnarray}\label{833}
 \sum_{\beta}\lambda^{A}_{\beta}(\boldsymbol{p_{z}})\stackrel{\neg}{\lambda}_{\beta}^{A}(\boldsymbol{p_{z}}) &=& -m(\mathbbm{1}-G_z),                                                                                                             
\end{eqnarray}
where $G_z$ is written as 
\begin{eqnarray}
G_z = \left(\begin{array}{cccc}
0 & 0 & 0 & -i \\ 
0 & 0 & i & 0 \\ 
0 & -i & 0 & 0 \\ 
i & 0 & 0 & 0
\end{array} \right).
\end{eqnarray}
Note that the spin sums presented in \eqref{83} and \eqref{833} hold invariant under any Lorentz transformation, for the sake of being momentum (coordinate) independent, in other words, this prescription clearly bypass the issue of Lorentz invariance. 

A parenthetic remark, the correct fashion to display the spin sums is $G_z(\boldsymbol{p})$, due to the fact that the spinors presented here are written in terms of the momentum $\boldsymbol{p}$, however, as performed in the previous literature, please check Ref \cite{hep-ph}, even such term does not explicitly depends on $\boldsymbol{p}$ and it is only $\phi$ dependent, then, authors decided to write it as $G(\phi)$. Following the same akin reasoning, in the context presented in this manuscript, we have no explicit dependence on $\boldsymbol{p}$ or $\phi$, so, under such circumstances, it is more convenient to translate it into $G_z$, looking towards avoid any kind of confusion.

\section{Phenomenological approach: The Yukawa and the Self-Interaction}\label{phenom}
This section is devoted to explore the Elko \emph{TYPE 1} spinors phenomenological features. In such a way, in the context developed up until now, we will analyse the cross-section produced by a particle restricted to move along a line, between an initial state, $|i\rangle$, and a final state, $|f \rangle$. The first step lies in the computation of the $\mathcal{S}^2$-Matrix \cite{landau, peskin}, such amount will describe the transition probability among the mentioned quantum states and it is given by
\begin{eqnarray}
\lvert \mathcal{S}_{fi}\rvert^2=\lvert \langle f|\hat{\mathcal{S}}|i\rangle \rvert ^2,
\label{s}
\end{eqnarray}
where $\hat{\mathcal{S}}=e^{i\hat{H}t}$ stands for the temporal evolution operator and it must be unitary.
We will consider a two Elko's scattering process, e.g., $\mathrm{f}_{\alpha}({\bf p})+\mathrm{f}_{\alpha'}({\bf p'})\rightarrow \mathrm{f}_{\beta}({\bf k})+\mathrm{f}_{\beta'}({\bf k'})$, governed by the Yukawa interaction $\mathcal{L}_{int} = \varepsilon_{E}\stackrel{\neg}{\mathrm{f}}(x)\mathrm{f}(x)\phi(x)$ \cite{jcap}, where $\mathrm{f}$ stand for the Elko quantum field. The creation and annihilation operators are given in \cite{jcap,AH1}. The dimension of the coupling constant is $[\varepsilon_{E}]=[\text{mass}]$ and the lower indexes $\alpha,\alpha',\beta$ and $\beta'$ stands for the helicity $\{\pm,\mp\}$. 
Here we will work only to the lowest order, so, such process can be represented by the diagrams 
\begin{figure}[h!]
\centering
\includegraphics[scale=0.60]{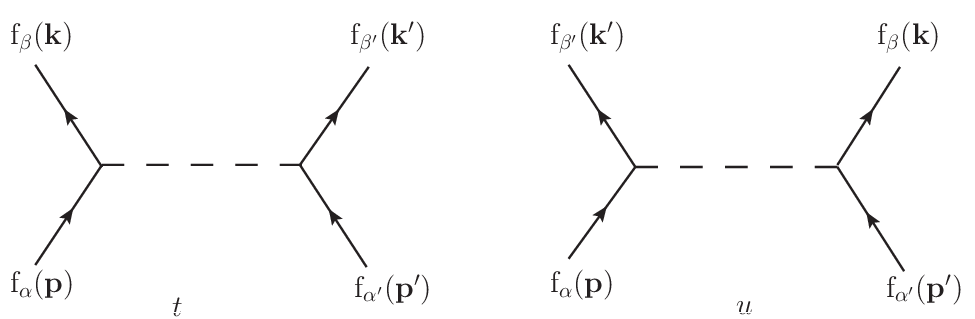}
\caption{Elko scattering mediated by a real scalar boson in the $t$ and $u$ channels.}
\label{sc}
\end{figure}

Moreover, it is important to highlight that we are in an unidimensional framework, in other words, the elastic scattering occurs in the center of mass along to the preferred axis, accordingly to \cite{vassalo}. Defining
\begin{eqnarray}
p^{\mu}&=&\left(\frac{E}{2},0,0,p_{z}\right), \qquad p'^{\mu}=\left(\frac{E}{2},0,0,-p_{z}\right), \\ \nonumber
k^{\mu}&=&\left(\frac{E}{2},0,0,-k_{z}\right), \qquad k'^{\mu}=\left(\frac{E}{2},0,0,k_{z}\right), \\ \nonumber
\end{eqnarray}
where ${\bf p}=p_{z}\hat{z}$, ${\bf p'}=-p_{z}\hat{z}$, ${\bf k}=-k_{z}\hat{z}$ and ${\bf k'}=k_{z}\hat{z}$, thus, the $\mathcal{S}_{fi}$ matrix reads
\begin{eqnarray}\label{666}
\mathcal{S}_{fi}=i(2\pi)^4\delta^{4}(p_{i}-p_{f})\mathcal{M}_{fi},
\end{eqnarray}
so, we have
\begin{eqnarray}\label{488}
\mathcal{M}_{(\mathrm{f}_{\beta}\mathrm{f}_{\beta'})(\mathrm{f}_{\alpha}\mathrm{f}_{\alpha'})}= 
\frac{\varepsilon_{E}^2}{m^2}\Bigg[\left(\stackrel{\neg}{\lambda}_{\beta}^{S}({\bf k})\lambda_{\alpha}^{S}({\bf p})\right)\frac{1}{q^2-m_{\phi}^2}\left(\stackrel{\neg}{\lambda}_{\beta'}^{S}({\bf k'})\lambda_{\alpha'}^{S}({\bf p'})\right)
 -\left(\stackrel{\neg}{\lambda}_{\beta'}^{S}({\bf k'})\lambda_{\alpha}^{S}({\bf p})\right)\frac{1}{r^2-m_{\phi}^2}\left(\stackrel{\neg}{\lambda}_{\beta}^{S}({\bf k})\lambda_{\alpha'}^{S}({\bf p'})\right)\Bigg],\nonumber\\ 
\label{m1}
\end{eqnarray}
where $q^{\mu}=(k^{\mu}-p^{\mu})$ and $r^{\mu}=(k'^{\mu}-p^{\mu})$ stands for the Higgs boson four-momentum vector and $m_{\phi}$ stands for it mass. Taking into account the incident and the scattered particles, after a straightforward calculations we obtain
\begin{eqnarray}
&&\rvert \overline{\mathcal{M}}_{(\mathrm{f}_{\beta}\mathrm{f}_{\beta'})(\mathrm{f}_{\alpha}\mathrm{f}_{\alpha'})}\lvert^{2} = 16\varepsilon_{E}^4\times\left(\frac{1}{(t-m_{\phi}^2)^2}+\frac{1}{(u-m_{\phi}^2)^2}-\frac{1}{(t-m_{\phi}^2)(u-m_{\phi}^2)}\right),
\end{eqnarray}
where $t=(k^{\mu}-p^{\mu})$ and $u=(k'^{\mu}-p^{\mu})$ are the well-known Mandelstam variables. The result obtained clearly evince that along the preferred axis the Yukawa interaction hold unitary due the convergence attribute in the high energy limit encoded in the Mandelstam variables.

A more comprehensive study of the interaction with the Higgs boson and the possible Elko production channels at particle colliders can be seen in \cite{LHC, mono}. Here, we are only interested in analyse the Elko scattering probability in one dimension by confronting the results for the same processes in $(3+1)$-dimensions existing in the literature.
As one know from the Elko literature, in a tree level and in absence of a preferred direction, as shown in \cite{elko5}, self-interaction do not preserve the unitarity. Now, the next task is to analyse the process formed by $\mathrm{f}_{\alpha}({\bf p})+\mathrm{f}_{\alpha'}({\bf p'})\rightarrow \mathrm{f}_{\beta}({\bf k})+\mathrm{f}_{\beta'}({\bf k'})$, along $z$-axis, described by a self-interaction $\mathcal{L}_{int} = g_{a}(\stackrel{\neg}{\mathrm{f}(x)}\mathrm{f}(x))^2$ where $g_{a}$ play a role of a dimensionless coupling constant.

\begin{figure}[h!]
\centering
\includegraphics[scale=1.5]{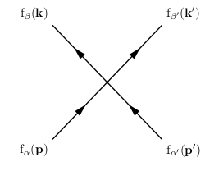}
\includegraphics[scale=1.5]{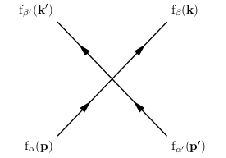}
\caption{Elko self-interaction.}
\label{el}
\end{figure}

The Feynman diagrams for this case are displayed in Figure \eqref{el}

The transition amplitude for this process is given by
\begin{eqnarray}
\mathcal{M}_{(\mathrm{f}_{\beta}\mathrm{f}_{\beta'})(\mathrm{f}_{\alpha}\mathrm{f}_{\alpha'})}=
\frac{g_{a}}{m^2}\Bigg[\left(\stackrel{\neg}{\lambda}_{\beta}^{S}({\bf k})\lambda_{\alpha}^{S}({\bf p})\right)\left(\stackrel{\neg}{\lambda}_{\beta'}^{S}({\bf k'})\lambda_{\alpha'}^{S}({\bf p'})\right) 
 -\left(\stackrel{\neg}{\lambda}_{\beta'}^{S}({\bf k'})\lambda_{\alpha}^{S}({\bf p})\right)\left(\stackrel{\neg}{\lambda}_{\beta}^{S}({\bf k})\lambda_{\alpha'}^{S}({\bf p'})\right)\Bigg].
\label{mel}
\end{eqnarray}
We evaluate the squared unpolarized amplitude, given by $\rvert\overline{\mathcal{M}}_{(\mathrm{f}_{\beta}\mathrm{f}_{\beta'})(\mathrm{f}_{\alpha}\mathrm{f}_{\alpha'})}\lvert^{2}$, in order to ascertain the unitarity
and after some procedures we obtain
\begin{eqnarray}
\rvert\overline{\mathcal{M}}_{(\mathrm{f}_{\beta}\mathrm{f}_{\beta'})(\mathrm{f}_{\alpha}\mathrm{f}_{\alpha'})}\lvert^{2} = 16 g_{a}^2,
\end{eqnarray}
where we note that the last result is momentum and mass independent.

Thus, the results for scattering amplitudes in the interactions studied, have no dependence on the azimuthal angle $\phi$ and are not proportional to the momentum and the energy of the center of mass. Such features comes from the spin sums for the Elko field, which along the privileged axis, is independent of these quantities and so,
as it is imposed by QFT, we found that in the first order of perturbation, the probability amplitude for two possible Elko interactions is conserved in the preferred direction. 

It is a well known fact that the cross-section, $\sigma$, is a measurable quantity which can be obtained from the scattering amplitude computation, $\sigma \propto |\mathcal{M}_{fi}|^2$ \cite{peskin,hovakimian}. 
Since both amplitudes evaluated in this communication converge in the high energy regime (limit), the tree unitarity of the interactions analysed along $z$-axis has strong indications of being guaranteed \cite{tree}.

On the other hand, the unitarity of $\mathcal{S}$-matrix is incorporated at all perturbative levels if the optical theorem is satisfied, which in general requires the computation of loops in order to extract the respective imaginary
parts. As we have seen that in the first order of perturbations, the couplings produce unitary amplitudes and that this, in turn, is the greater contribution to the scattering studied, the calculations in higher orders should not provide a divergence of these processes at high energies.

The generalization of the mentioned theorem for a theory in $N$ spatial dimensions is shown in \cite{boya,barlette}, showing and reinforcing the possibility to find the unitarity in any dimension, including analysis in lower dimensions such as our case ${N=1}$. This is a fair motivation to study phenomenologies associated with invariant theories within VSR context, which can retrieve the theory along some preferred direction.

\section{Final Remarks}\label{conclusions}
Although, recently a new set of mass-dimension-one \emph{new local fields} were proposed, we choose to analyse the Elko spinors in its first  formulation. As already mentioned, both formulations are relevant and, although the recent formulation exhibits an invariant spin sums under Lorentz transformations whilst the previous one shows invariance by the VSR theory, they are part of different contexts. Thus, more than an academic exercise, we built a useful set of Elko spinors defined in terms of the momentum in Cartesian coordinates system ready to be used for phenomenology scholars that, as noted, it is still lacking in the current literature besides being something useful. Following the paper schedule, we investigated all the properties of Elko along the local axis, in such a way that we can claim the theory as being unitary, in given framework we find the hidden unitarity, searching for a prominent physics.  In this proposed scenario, we computed the spin sums and in this context such physical amount manifest Lorentz invariant, entailing a unitary theory. Through phenomenological analysis, it was verified that Yukawa interaction is convergent even in a high energy limit and the self-interaction, besides convergent, depending on coupling constant.      

\section{Acknowledgement}
LCD, RdCL, RJBR and CHCV are grateful to Professor José Abdalla Helayël-Neto for the appreciation, helpful discussions and suggestions on the original manuscript during its writing stage, authors also thanks to Eslley Scatena for the privilege of his revision, comments and appreciation of this work and thanks to Dino Beghetto for discussions and advices on this essay. Authors also thanks the Referees, their questions helped to substantially improve the manuscript. LCD and RdCL thank to CAPES, RJBR thanks to CNPq (Grant Number 155675/2018-4) and CHCV thanks CNPq (Grant Number 300236/2019-0) for the financial support.

\end{document}